\documentclass[superscriptaddress,reprint,amsmath,twocolumn,amssymb,prx, nodate]{revtex4-2}

\usepackage{graphicx} 
\usepackage{amsmath} 
\usepackage{physics}
\usepackage{mathtools}
\usepackage{enumitem}
\usepackage[ruled]{algorithm2e}
\usepackage{booktabs}

\usepackage[version=4]{mhchem}

\usepackage{hyperref} 
\usepackage[noabbrev]{cleveref} 
\crefformat{equation}{Eq.~(#2#1#3)}
\crefformat{figure}{Fig.~#2#1#3}
\crefformat{table}{Table #2#1#3}

\usepackage[load=prefixed,load=abbr,separate-uncertainty=true]{siunitx} 

\newcommand{\sub}[1]{_{\mathrm{#1}}}

\newcommand{\ie}{i.e.\@}

\newcommand{\bigO}[1]{\mathcal{O}(#1)}

\usepackage{color, colortbl}
\definecolor{Gray}{gray}{0.9}

\newcommand{\etal}{\emph{et~al.}}

\newcommand{\Ham}{\mathcal{H}}

\newcommand{\No}{N_{\mathrm{O}}}
\newcommand{\Nq}{\No}

\renewcommand{\vec}[1]{\vb{#1}}

\newcommand{\sig}[2]{\sigma^{#1}_{#2}}

\usepackage{xparse}

\NewDocumentCommand\se{gg}{%
	\ensuremath{
		x\IfNoValueTF{#1}{}{_{#1}}\IfNoValueTF{#2}{}{^{#2}}
	}
}

\NewDocumentCommand\s{gg}{%
	\ensuremath{
		\vec{x}\IfNoValueTF{#1}{}{_{#1}}\IfNoValueTF{#2}{}{^{#2}}
	}
}

\begin{document}

\title{Autoregressive neural-network wavefunctions for \emph{ab initio} quantum chemistry}

\author{Thomas D.\ Barrett}
\email{t.barrett@instadeep.com}
\affiliation{InstaDeep, London, W2 6LG, UK}
\affiliation{University of Oxford, Clarendon Laboratory, Parks Road, Oxford,  OX1 3PU, UK}

\author{Aleksei Malyshev}
\affiliation{University of Oxford, Clarendon Laboratory, Parks Road, Oxford,  OX1 3PU, UK}

\author{A.\ I.\ Lvovsky}
\email{alex.lvovsky@physics.ox.ac.uk}
\affiliation{University of Oxford, Clarendon Laboratory, Parks Road, Oxford,  OX1 3PU, UK}
\affiliation{Russian Quantum Center, Skolkovo, 143025, Moscow, Russia}



\begin{abstract}

In recent years, neural network quantum states (NNQS) have emerged as powerful tools for the study of quantum many-body systems. 
Electronic structure calculations are one such canonical many-body problem that have attracted significant research efforts spanning multiple decades, whilst only recently being attempted with NNQS.  However, the complex non-local interactions and high sample complexity are significant challenges that call for bespoke solutions.
Here, we parameterise the electronic wavefunction with a novel autoregressive neural network (ARN) that permits highly efficient and scalable sampling, whilst also embedding physical priors reflecting the structure of molecular systems without sacrificing expressibility.
This allows us to perform electronic structure calculations on molecules with up to 30 spin-orbitals -- at least an order of magnitude more Slater determinants than previous applications of conventional NNQS -- and we find that our ansatz can outperform the de-facto gold-standard coupled cluster methods even in the presence of strong quantum correlations.
With a highly expressive neural network for which sampling is no longer a computational bottleneck, we conclude that the barriers to further scaling are not associated with the wavefunction ansatz itself, but rather are inherent to any variational Monte Carlo approach.
\end{abstract}

\maketitle

\section{Introduction}

The grand challenge of \emph{ab initio} quantum chemistry (QC) is to solve the many-body Schr\"{o}dinger equation describing interaction of heavy nuclei and orbiting electrons.  In principle, such solutions can provide complete access to a molecular entity's chemical properties, however, in practice, the remarkable complexity of these systems makes them intractable for all but the simplest of cases.  Indeed, the many-body electronic problem is fundamentally NP-hard~\cite{qchem_np_complete_1, qchem_np_complete_2}, which has motivated significant efforts to tailor numerical methods to QC systems~\cite{hammond94,langhoff12,piela13}.

The QC task we pursue here is to find the ground state of a molecular system containing $N_{\mathrm{e}}$ electrons.  Under the Born-Oppenheimer approximation the molecular Hamiltonian 
 can be written (using atomic units) in the position basis as~\cite{mcardle20}
\begin{equation}
		\mathcal{H} = 
			- \sum_{i} \frac{\nabla_{i}^{2}}{2}
			- \sum_{i,I} \frac{Z_I}{\abs{\vb{r}_i - \vb{R}_I}}
			+ \frac{1}{2} \sum_{i \neq j} \frac{1}{\abs{\vb{r}_i - \vb{r}_j}}.
\label{eq:TISE}
\end{equation}
with $\vb{r}_i$ being the spatial position of the $i$-th electron, $\vb{R}_I$ and $Z_I$ denoting the position and atomic number of the $I$-th nucleus. The state of interest satisfies the time-independent Schr\"odinger equation 
\begin{equation}
    \mathcal{H}\ket\psi=E\ket\psi
\end{equation}
and its wavefunction $\langle\vb{\xi}_1,\dots,\vb{\xi}_{N_{\mathrm{e}}}|\psi\rangle=\psi(\vb{\xi}_1,\dots,\vb{\xi}_{N_{\mathrm{e}}})$ is represented in coordinates $\vb{\xi}_i {=} (m_{s,i}, \vb{r}_i)$, which include both the spin projection and spatial position of the electron. 
Importantly, the fermionic nature of electrons means that the wavefunction must obey antisymmetric exchange symmetries, i.e.\@ $\psi(\dots,\vb{\xi}_i,\dots,\vb{\xi}_j,\dots) {=} {-}\psi(\dots,\vb{\xi}_j,\dots,\vb{\xi}_i,\dots)$.

QC methods typically define a basis set of $N_{\mathrm{O}} > N_{\mathrm{e}}$ single-electron spin-orbitals, $\{\chi_i(\vb{\xi})\}_{i=1,\dots,N_{\mathrm{O}}}$ and represent the electronic state as a linear combination of the form 
\begin{equation}\label{eq:state}
\ket\psi=\sum_k \psi_k \ket*{\vb{x}_{k}},    
\end{equation}
where each of the components $\ket*{\vb{x}_{k}}$ is uniquely identified by the subset of basis spin-orbitals that are occupied by an electron. In the second quantization,  these components are written as occupation vectors,
\begin{equation}\label{eq:Slater}
  \ket*{\vb{x}_{k}} {\equiv} \ket*{x_{k}^{1}, \dots, x_{k}^{N_{\mathrm{O}}}},
\end{equation}
where $x_{k}^{j} \in \{0,1\}$ denotes whether the $j$-th spin-orbital of the $k$-th component is occupied, with $\sum_{j=1}^{N_{\mathrm{O}}}x_{k}^{j}=N_{\mathrm{e}}\ \forall\ k$. In the first quantization, these components are represented by antisymmetric tensor products of the $N_{\mathrm{e}}$ occupied orbitals~\cite{mcardle20}, hence we hereafter refer to them as ``(Slater) determinants" and the corresponding sequences $\vb{x}_{k}$ as ``configuration strings".

As the number of possible determinants grows exponentially with the system size, so-called ``full configuration interaction'' (FCI) methods --- those considering the entire space of Slater determinants --- rapidly become intractable for larger systems.

To combat this scaling issue, leading \emph{ab-initio} QC approaches often consider systematic corrections to a reference state (typically the Hartree-Fock state, the single lowest energy Slater determinant).  Configuration interaction (CI) methods~\cite{sherrill99} restrict the active space to configuration strings that are different from the reference by no more than a certain number of excitations.
By contrast, coupled cluster (CC) approaches~\cite{coester60, bartlett07} can access arbitrarily excited Slater determinants using non-linear combinations of excitation operators up to a certain order, but cannot parameterise arbitrary superpositions (i.e.\@ do not offer fully general $\psi_k$'s in \cref{eq:state}).
Whilst both methods trade off expressibility for reduced complexity, they are specifically designed for the typical structure of molecular wavefunctions and often provide good performance.  However, even the more accurate CC approach -- which is often considered the leading \emph{ab-initio} method -- still may fail in the presence of strong static correlations (i.e.\@ when the wavefunction can no longer be adequately described by systematic corrections to a single Slater determinant)~\cite{ccsdt_failure_1, bartlett07}.
An alternative is to use compact parameterisations of the electronic wavefunction --- such as the Jastrow-Slater ansatz~\cite{foulkes01} or matrix-product states~\cite{white92,white99} --- which can then be optimised to find the ground state using stochastic~\cite{nightingale98} or non-stochastic~\cite{neuscamman11} methods.

The limitations of these approaches are typically determined by the expressiveness of the wavefunction ansatz and the efficiency with which it can be optimised, and it is here that machine learning (ML) offers great potential.
The approach is based on the assumption that the ground state is dominated by relatively few ($K\ll{N_{\mathrm{O}} \choose N_{\mathrm{e}}}$) Slater determinants.  A generative neural network, with parameters $\theta$, can encode the molecular wavefunction and produce \emph{samples} of Slater determinants $\vb{x}_{k}$ with the (\emph{a priori} unknown) probabilities $|\psi_{\theta}(\vb{x}_{k})|^2$.
A batch of $N$ (where $K\ll N\ll{N_{\mathrm{O}} \choose N_{\mathrm{e}}}$) samples, with the associated phases typically determined by an auxiliary network, then constitutes a reasonably accurate approximation of that wavefunction. Standard iterative methods can then optimize the network parameters to yield the lowest energy expectation value.

In 2017, Carleo and Troyer demonstrated that neural networks -- specifically restricted Boltzmann machines (RBMs) -- can parameterise a many-body wavefunction and capture non-trivial correlations within the exponentially large encoded Hilbert space~\cite{carleo17}.  Subsequently, neural network quantum states (NNQS) have proven to be successful variational ansatz for problems such as finding the ground state of interacting spin-systems~\cite{sharir20, allah20}, quantum state tomography~\cite{torlai18, carrasquilla19, neugebauer20, ahmed20} and classical simulations of quantum computing~\cite{jonsson18}.  Whilst most of the development of NNQS has been within the context of condensed matter physics (CMP), recently (2020) Choo \emph{et~al.}~\cite{choo20} demonstrated that the fermionic electronic structure problem can be mapped to an equivalent optimisation problem on a system of interacting spins, opening the door to applying NNQS to QC.  However, RBM models rely on inherently inefficient sampling procedures --- such as Markov Chain Monte Carlo (MCMC) methods --- to approximate the encoded Boltzmann distribution during training. 
Moreover, RBMs are ``black-boxes'' that do not reflect our significant physical knowledge of molecular systems.
Therefore, the search for alternative neural network architectures that  overcome these shortcomings remains an important item on the agenda of this developing field.

In this work we present a novel neural network ansatz for second-quantised molecular wavefunctions that addresses these outstanding challenges.  By considering an autoregressive decomposition of the wavefunction, we demonstrate a highly efficient sampling algorithm, which is inherently parallelised and scales with the number of unique configuration strings sampled instead of the overall sample batch size.  The feedforward architecture used allows us to train the network using standard backpropagation techniques \cite{lecun88} and, moreover, embed important physical priors (i.e.\@ conserved quantities and invariances) into the wavefunction without sacrificing expressibility.  Ultimately, this allows us to approach the true FCI solution on systems at least 25 times larger than what was previously achieved using a conventional NNQS.
More broadly, our work represents an ambitious application of NNQS to a challenging system where conventional approaches can fail, which is an important milestone for the field as it progresses from its nascent promise and theoretical investigations~\cite{gao17}, to tackling a broader range of problems in quantum many-body physics~\cite{choo19}.  

\section{Neural network wavefunction}
\label{sec:neural_network_wavefunction}

\subsection{Variational Monte-Carlo optimization}
\label{sec:vmc_opt}

We employ the Jordan-Wigner encoding~\cite{wigner28} which treats the state \eqref{eq:state} as an ensemble of interacting qubits. The Hamiltonian \eqref{eq:TISE} then takes the form  (see Methods for details)
\begin{equation}
	\mathcal{H}_{\mathrm{Q}} = \sum_{j} h_j \prod_{i=1}^{N_{\mathrm{O}}} \sigma^{\nu_{j,i}}_{i},
	\label{eq:HamSpin}
\end{equation}
where $\sigma^{\nu_{j,i}}_{i}$ is a Pauli operator acting on the $i$-th qubit ($\nu_{j,i} \in \{ \mathrm{I}, x, y, z \}$).
Whilst such a model is often considered as a system of interacting spins in CMP, we will henceforth refer to qubits and reserve ``spin'' exclusively for the spin-state of electrons to avoid confusion.

Let the component amplitudes of the state \eqref{eq:state} depend on a set of ansatz parameters, $\theta$, such that
\begin{equation}
	\ket{\psi_{\theta}} = \sum_{k} \psi_{\theta}(\vb{x}_{k}) \ket*{\vb{x}_{k}}.
\end{equation}
This wavefunction can be optimized (trained) to find the ground state of the electronic system using a variational Monte-Carlo (VMC) approach~\cite{choo20}.  The energy expectation value is given by
\begin{equation}
	E
	=  \ev{\mathcal{H}_{\mathrm{Q}}}{\psi_\theta} 
	= \sum_{k} E_{\mathrm{loc}}(\vb{x}_{k}) \abs{\psi_{\theta}(\vb{x}_{k}) }^{2},
	\label{eq:energy}
\end{equation}
where we have defined the ``local energy" of a given Slater determinant as
\begin{equation}
	 E_{\mathrm{loc}}(\vb{x}_{k}) = \sum_{k^{\prime}} \frac{ \psi_{\theta}(\vb{x}_{k^{\prime}}) }{ \psi_{\theta}( \vb{x}_{k}) } \mel*{\vb{x}_{k}}{\mathcal{H}_{\mathrm{Q}}}{\vb{x}_{k^{\prime}}}.
	\label{eq:Eloc}
\end{equation}
and assumed that the normalization $\braket{\psi_\theta}=1$ is enforced by the ansatz, which is the case for the ARN we implement.

An unbiased estimate of the energy \eqref{eq:energy} is given by the mean local energy across a batch of samples drawn from the underlying distribution defined by the wavefunction $p( \vb{x}_{k} ) \equiv \abs{\psi_\theta(\vb{x}_{k})}^{2}$.  These samples can similarly estimate the gradient of energy with respect to our variational parameters,
\begin{equation}
	\nabla_\theta E =
	2\Re \big( \mathbb{E}_{p} \left[ E_{\mathrm{loc}} \nabla_\theta \ln(\psi_\theta^{\ast}) \right] \big)
		\label{eq:VMCgrad}
\end{equation}
where, for brevity, we have dropped any explicit dependence on the state, $\vb{x}_{k}$.
Therefore, the wavefunction can be optimised towards the ground state using standard backpropagation techniques.

\subsection{Autoregressive quantum states}

\begin{figure*}[!htb]
\centering
\includegraphics[width=0.95\textwidth]{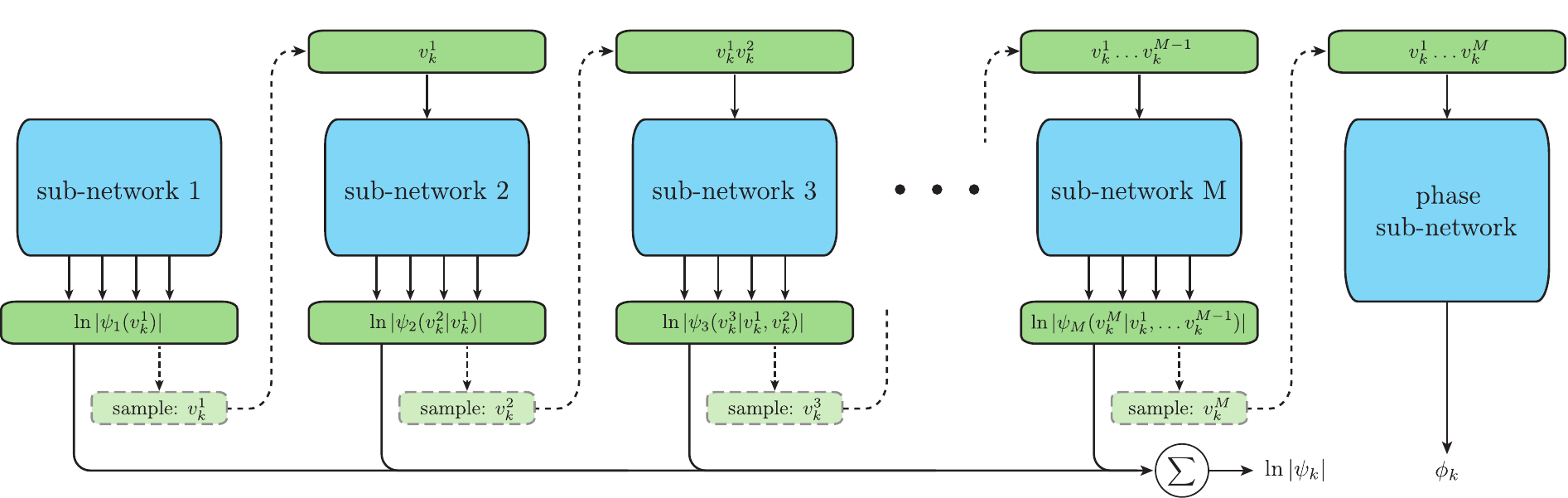}
\caption{The high level architecture of the ARN implementing our wavefunction ansatz.  Solid lines indicate the operations used for the inference of the absolute value $|\psi_k|$ of the Slater determinant $\ket{\vb{x}_{k}}$.  Blue blocks correspond to neural network operations, whose structure is further detailed in \cref{fig:subnetwork}.
}
\label{fig:network}
\end{figure*}

Our choice of wavefunction ansatz is an autoregressive neural network (ARN) -- a class of generative models originally developed within the ML community as tractable, feed-forward alternatives to RBMs~\cite{larochelle11, uria16} that have only very recently been demonstrated as viable NNQS in the context of condensed matter physics~\cite{sharir20, morawetz20}.
The basic principle of an ARN is to decompose the joint probability distribution across multiple random variables $z_i$ into a series of conditional distributions, e.g.\@ $\mathrm{pr}(z_1, \dots, z_n) = \prod_{i=1}^{n} \mathrm{pr}(z_i \vert z_1, \dots z_{i-1})$.
That is to say, instead of trying to model the distribution over every possible configuration of these discrete variables simultaneously, one instead considers each variable sequentially, with each subsequent distribution being a function of the variables that have come before.  

In our case, the distribution we wish to model is a wavefunction over all $N_{\mathrm{O}}$ qubits with complex coefficients $\psi_k = \mathrm{e}^{\mathrm{i} \phi_k}\abs{\psi_k}$.
As sampling a wavefunction only requires its absolute value we use an ARN only for $\abs{\psi_k}$.  The overall phase, $\phi_k$, is predicted separately by a standard feed-forward network that takes as input the entire configuration string.

The basis set of spin-orbitals used in our analysis consists of $M$ spatial orbitals, each of which is doubly represented for the upward and downward projections of the spin, i.e.\@ $N_{\mathrm{O}}{=}2M$ and $\ket*{\vb{x}_{k}} {\equiv} \ket*{x_{k}^{1\uparrow}, x_{k}^{1\downarrow}, \dots, x_{k}^{M\uparrow}, x_{k}^{M\downarrow}}$.  This leads to natural symmetries in the resulting wavefunction, as we discuss below, and so we treat each spatial orbital as a single unit, $v_k^i \equiv (x_{k}^{i\uparrow}, x_{k}^{i\downarrow})$, which can take on four possible configurations: $(0,0), (0,1), (1,0), (1,1)$.  Concretely, we consider wavefunction coefficients in the form
\begin{equation}
	\ln \psi_k = \sum_{i=1}^{M} \ln \abs*{\psi_i(v_k^i \vert v_k^1 \dots v_k^{i-1})} + \mathrm{i}  \phi(v_k^1\dots v_k^{M}).
	\label{eq:arWavefunction}
\end{equation}
So long as each conditional distribution is normalised, $\sum_{v_k^i \in\{(0,0), (0,1), (1,0), (1,1)\}} \abs*{ \psi_{i}(v_k^i \vert \dots ) }^{2} {=} 1$, the overall wavefunction is also normalised, $\sum_k \abs*{\psi_k}^{2} {=} 1$. 

\subsection{Inference and efficient sampling}
\label{sec:sampling}

\begin{algorithm}[b]
	\DontPrintSemicolon
	\small
	\SetKwInOut{Input}{input}\SetKwInOut{Output}{output}
	
	\Input{batch size, $N$.}
	\Output{unique samples, $\mathrm{X}$, their count, $\mathrm{N_X}$, and their log-probabilty amplitudes, $\ln\psi(\cdot)$}
	
	\Begin{
	
	\tcp*[l]{X : partially sampled configurations}
	\tcp*[l]{N$_\mathrm{X}$ : count of elements of X}
	\tcp*[l]{A : log prob.\ amp.\ for elements of X}

	initialise: $\mathrm{X}{=}\big[ [\,] \big]$, $\mathrm{N_X}{=}\big[N\big]$, $\mathrm{A}{=}\big[0\big]$
	
	\For{\textup{each orbital} $i=1,\dots,M$}{
		\tcp*[l]{Orbitals are processed sequentially.}
		initialise: $\mathrm{X}^{i}{=}[]$, $\mathrm{N}_\mathrm{X}^{i}{=}[]$, $\mathrm{A}^{i}{=}[]$\;
	
		\For{$k=1,\dots,\mathrm{len}(\mathrm{X})$}{
		\tcp*[l]{Samples are processed in parallel.}
			$[v_{k}^{1}, \dots, v_{k}^{i-1}],\ n_k,\, \ln\psi_k = \mathrm{X}[k],\ \mathrm{N}_\mathrm{X}[k],\, \mathrm{A}[k]$\;
			sub-network $i$ forward pass: $\ln\abs*{ \psi_{i}( \cdot \vert v_{k}^{1} {\dots} v_{k}^{i{-}1} )}$\;
			take $n_k$ samples:  $v_{k}^{i} \sim \abs*{ \psi_{i} ( \cdot \vert v_{k}^{1} \dots v_k^{i-1} ) }^{2}$\;
			\For{$v_{k}^{i}$\ \textup{sampled}\ $n_k^{i}$ \textup{times}}{
				\If{$n_k^{i} > 0$}{
				$\mathrm{X}^i.\mathrm{append}([v_{k}^{1}, \dots, v_{k}^{i-1}, v_{k}^{i}])$\;
				$\mathrm{N}_\mathrm{X}^{i}.\mathrm{append}(n_k^{i})$\;
				$\mathrm{A}^{i}.\mathrm{append}( \ln\abs*{\psi_k} {+} \ln\abs*{\psi_{i}( v_{k}^{i} \vert v_{k}^{1} \dots v_{k}^{i-1} ) )}$
				} 
			} 
		} 
		$\mathrm{X} \leftarrow \mathrm{X}^i$\;
		$\mathrm{N_X} \leftarrow \mathrm{N}_\mathrm{X}^i$\;
		$\mathrm{A} \leftarrow \mathrm{A}^i$\;
	} 
	
	$\ln\psi(\mathrm{X}) = \mathrm{A} + \mathrm{i}\phi(\mathrm{X})$\;
	
	\Return $\mathrm{X}$, $\mathrm{N_X}$, $\ln\psi(\mathrm{X})$\;
	} 
	\caption{Batched sampling procedure
	\label{alg:sampling}} 
\end{algorithm}

Our network architecture is shown in \cref{fig:network}. The autoregressive property is realised by setting up $M$ sub-networks, whose structure is shown in \cref{fig:subnetwork}. The $i$-th sub-network takes the partial configuration string $(v_k^1 \dots v_k^{i-1})$ as input and outputs the normalised log-amplitudes $\ln \abs*{ \psi_i(v_k^i \vert v_k^1 \dots v_k^{i-1}) }$ of the four possible configurations of $v_k^i$. The network can be used in two different modes which we refer to as `inference' and `sampling'.  Inference refers to the task of evaluating the log-amplitudes of a given configuration string, $\vb{x}_{k}$, and corresponds to a single forward pass of the network.  In this setting, all sub-networks can run in parallel and their outputs are added according to \cref{eq:arWavefunction}.

The task of sampling is to generate configuration strings $\vb{x}_{k}$ according to the underlying distribution $\abs{\psi}^{2}$. As we show below, our method is inherently tailored to output \emph{unique} configuration strings. Each such string is associated with a number $n_k$ indicating how many times it would occur if the standard procedure, consisting of independently sampling the configuration $N$ times, were employed. The number $N=\sum_kn_k$ is hereafter referred to as the batch size. 

The procedure is detailed in Algorithm \ref{alg:sampling}. As we sequentially progress through the orbitals, we maintain the information in three datasets: unique samples of partial configuration strings $(v_{k}^{1}, \dots, v_{k}^{i-1})$, their ``number of occurrences"  $n_k$, and the corresponding log-amplitudes $\ln\psi_k$, where $i$ is the current orbital number and $k$ indexes the unique samples.  We  supply each $(v_{k}^{1}, \dots, v_{k}^{i-1})$ to the $i$-th sub-network, which then yields the probabilities $\ln\abs*{ \psi_{i}(v_k^i  \vert v_{k}^{1} {\dots} v_{k}^{i{-}1} )}$\ for each $v_k^i \in\{(0,0), (0,1), (1,0), (1,1)\}$. We then use these probabilities to sample the number of occurrences for each $v_k^i$ from a multinomial distribution with the total number of trials equal to $n_k$.

We see that in contrast to standard methods, which sample configuration strings (and hence can produce massively redundant results), our method samples \emph{numbers of occurrences}, $n_k$, for each unique string. Importantly, a string is added to the dataset for each subsequent orbital only if the corresponding $n_k$ is non-zero. This prevents exponential growth of the dataset size as we progress through the orbitals.

This sampling procedure has multiple highly desirable properties.
\begin{enumerate}[label=(\roman*)]
\item The computational cost of generating both a sample and its associated wavefunction coefficient is the same as a single forward pass of the full network during inference.
\item Different inputs $(v_{k}^{1}, \dots, v_{k}^{i-1})$ to the $i$-th subnetwork can be processed in parallel.
\item The practical cost of generating a batch of samples scales with the number of unique configurations sampled, as opposed to the overall sample batch size.
\end{enumerate}
Point (\romannumeral 3) is especially beneficial when sampling highly asymmetric systems --- those where a single or few basis elements dominate the underlying distributions.  This is exactly the case in QC where the Hartree-Fock state typically dominates the low energy wavefunctions and will therefore be repeatedly sampled.  Therefore, obtaining meaningful statistics with which to optimise the wavefunction to below the Hartree-Fock energy requires massive numbers of samples where each batch will typically contain orders of magnitude fewer unique states.

Properties (\romannumeral1)-(\romannumeral3) are a significant departure from the standard 
MCMC~\cite{brooks11} algorithms commonly used to sample variationally optimised ansatz~\cite{carleo17}.
These approaches rely on multiple sequential evaluations of proposed configurations, with the computational cost increasing with the required sample set quality and the total number of samples.
When applying RBMs to QC using a Metropolis sampling scheme~\cite{hastings70}, Choo~\emph{et~al.}~\cite{choo20} observed that performance depended strongly on the number of samples used, but were limited to maximum batch sizes of $N=10^{6}$.  By contrast, our autoregressive ansatz readily generates sample sets with equivalent batch sizes of up to $N=10^{12}$.  Indeed, even larger batch sizes could be used but we found that either this did not improve performance or, as we will discuss, calculating the local energies for the increasing number of unique states generated became too computationally expensive.

\subsection{Encoding physical constraints}

We treat the molecule as a system of $N_{\mathrm{O}}$ qubits, with the corresponding Hilbert space dimension being $2^{N_{\mathrm{O}}}$.
A naive NNQS ansatz will predict non-zero probability for each of the $2^{N_{\mathrm{O}}}$ basis configurations $(x_{k}^{1}, \dots, x_{k}^{N_{\mathrm{O}}})$. However, only a small fraction of these configurations correspond to physically viable Slater determinants, particularly because the number of occupied spin-orbitals must equal the number $N_{\mathrm{e}}$ of electrons. Additional constraints arise from known physical symmetries and conservation laws.

Rather than having to learn these constraints during training, it is desirable that an NNQS incorporate them in an \emph{a priori} fashion.  By keeping focused on a relatively small, physically meaningful subset of possible outputs, such a network helps streamline the training and reduce the computational costs.

\begin{figure}[!tb]
\centering
\includegraphics{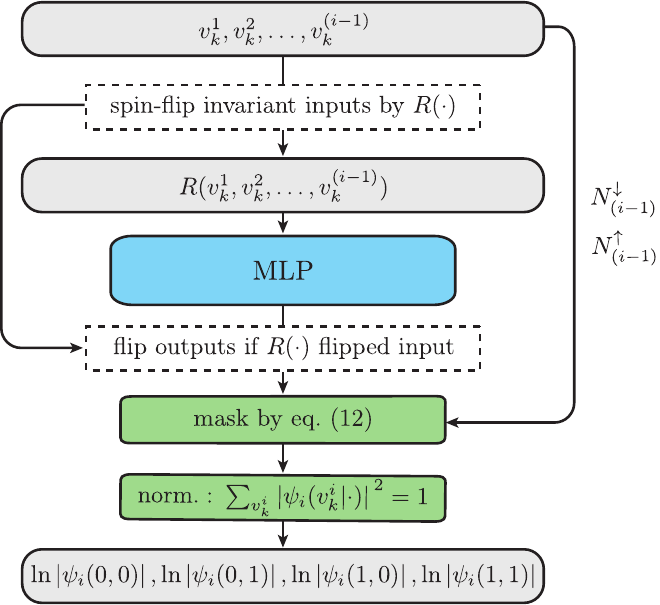}
\caption{The operation of a single conditional wavefunction sub-network. A multi-level perceptron (MLP) is the core of the sub-network, whose inputs and outputs are pre-and post-processes to enforce the physical prior: the conditional spin-flip $R(\cdot)$ ensures spin-flip symmetry as per \cref{eq:R}, whilst the mask eliminates unphysical configurations that do not conserve the electron number and multiplicity.
}
\label{fig:subnetwork}
\end{figure}

Our autoregressive architecture satisfies this requirement, without sacrificing expressibility, by using the sub-network structure shown in \cref{fig:subnetwork}, which combines a trainable multi-level perceptron (MLP) with hard-coded pre-and post-processing that enforces the priors.  This capability is a further important advantage of ARNs with respect to other NNQS ansatz such as RBMs, which  must instead rely on modified sampling methods to generate only physically viable states~\cite{choo20}. 

\subsubsection{Conservation of electron number and multiplicity}
\label{sec:masking}

The electron number and multiplicity of a molecule can be expressed as conserved quantities  -- $N_{\mathrm{e}}$ and $2S+1$, respectively -- both of which are known \emph{a priori}.  To utilize this information in our ARN, we first observe
that we only need to consider determinants with  the magnetic quantum number $\abs{M_S} \leq S$.  Moreover, due to the central symmetry of the problem, the energy of the molecule with a given total spin $S$ does not depend on $M_S$, so it is sufficient to restrict ourselves to any single valid $M_S$~\cite{sherrill95}.  
In this work we choose $M_S{=}S$.

With this assumption, we can write for the total number of occupied spin-up, $N_M^{\uparrow}$, and spin-down, $N_M^{\downarrow}$, orbitals:
\begin{equation}
	N_{\mathrm{e}} = N_M^{\uparrow} + N_M^{\downarrow},
	\quad
	 S= M_S = (N_M^{\uparrow} - N_M^{\downarrow})/2
	 \label{eq:conservationLaws}
\end{equation}
so $N_M^{\uparrow,\downarrow}=N_{\mathrm{e}}/2\pm S$.
In order for a partial configuration string $(x_{k}^{1\uparrow}, x_{k}^{1\downarrow}, \dots, x_{k}^{i\uparrow}, x_{k}^{i\downarrow})$ processed by a $i$-th ARN subnetwork to be consistent with the conditions (\ref{eq:conservationLaws}), it must satisfy the  requirements
\begin{equation}
\begin{aligned}
    &N_M^{\uparrow}-(M-i)\le N_i^{j\uparrow}\le N_M^{\uparrow}, \\
    &N_M^{\downarrow}-(M-i)\le N_i^{j\downarrow}\le N_M^{\downarrow},
    \label{eq:conservationLaws1}
\end{aligned}
\end{equation}
where
\begin{equation}
	N_i^{\uparrow} = \sum_{j=1}^{i} x_{k}^{j\uparrow}, \quad
	N_i^{\downarrow} = \sum_{j=1}^{i} x_{k}^{j\downarrow}.
\end{equation}
We mask each ARN subnetwork's outputs to assign a zero probability to any partial configuration string that does not satisfy \cref{eq:conservationLaws1}.

As a result of this simple modification, we achieve a remarkable reduction from the full $2^{N_{\mathrm{O}}}$ possible qubit configurations to ${M \choose N_M^{\uparrow}}{M \choose N_M^{\downarrow}}$.  We note that a similar technique was first used for NNQS by Hibat-Allah \emph{et~al.}~\cite{allah20} but, to date, has only been used to enforce zero magnetisation in the context of CMP~\cite{allah20, morawetz20}.

\subsubsection{Electron spin-flip symmetry}

The electronic Hamiltonian (\ref{eq:TISE}) does not explicitly depend upon electron spin.  Hence ``flipping'' every electron's spin should not change the absolute value of the amplitude associated with any specific Slater determinant.  Concretely, defining this flipping operation as
\begin{equation}
	F{:}
	\ket*{ x_{k}^{1\uparrow}, x_{k}^{1\downarrow}, \dots, x_{k}^{M\uparrow}, x_{k}^{M\downarrow} }
	{\rightarrow} 
	\ket*{x_{k}^{1\downarrow}, x_{k}^{1\uparrow}, \dots, x_{k}^{M\downarrow}, x_{k}^{M\uparrow} },
\end{equation}
we desire an ansatz such that ${ \abs*{\psi(\vb{x}_{k})} = \abs*{\psi(F(\vb{x}_{k}))} }$.

To achieve this, we enforce the equivalent symmetry onto \emph{each} sub-network, such that each conditional wavefunction satisfies
\begin{equation} \abs*{ \psi_{i}(v_{k}^{i} \vert v_{k}^{1} \dots v_{k}^{(i-1)}) } {=} \abs*{ \psi_{i}(F(v_{k}^{i}) \vert F( v_{k}^{1} \dots v_{k}^{i-1}) )}.
\label{eq:spinFlipSym}
\end{equation}
To that end, we pre-process the subnetwork input prior to entering it into the MLP, and post-process the MLP output as follows.

Recall that each sub-network has four outputs corresponding to the four possible configurations of ${v_{k}^{i} \in \{(0,0),(0,1),(1,0),(1,1)} \}$.  Applying $F(\cdot)$ to the network output is then simply swapping the predictions for the pair of singly-occupied states.  In the case that the input is spin-flip invariant ($F(v_{k}^{1} \dots v_{k}^{i-1}) = v_{k}^{1} \dots v_{k}^{i-1}$), \cref{eq:spinFlipSym} requires that log-amplitude of these singly-occupied states are equal.  This requirement is satisfied by assigning these log-amplitudes the same value at the post-processing stage.

To enforce the condition (\ref{eq:spinFlipSym}) for non-spin-symmetric inputs, we apply the following transformation to the subnetwork MLP input: 
\begin{align}\label{eq:R}
     &R(v_{k}^{1} \dots v_{k}^{i-1})\\\nonumber
     &=\left\{\begin{array}{l}(v_{k}^{1} \dots v_{k}^{i-1}) \ \textrm{if}\ n(v_{k}^{1} \dots v_{k}^{i-1})<n(F(v_{k}^{1} \dots v_{k}^{i-1}))\\
     F(v_{k}^{1} \dots v_{k}^{i-1}) \ \textrm{if}\ n(v_{k}^{1} \dots v_{k}^{i-1})>n(F(v_{k}^{1} \dots v_{k}^{i-1})),
     \end{array}
     \right. 
\end{align}
where $n(v_{k}^{1} \dots v_{k}^{i-1})$ is the binary number corresponding to the sequence $x_{k}^{1\uparrow}, x_{k}^{1\downarrow}, \dots, x_{k}^{i-1\uparrow}, x_{k}^{i-1\downarrow}$. In other words, $R(\cdot)$ maps a pair of spin-symmetric sub-network inputs to the same string, making the MLP output invariant to flipping the subnetwork input.  If the input configuration is such that it had to be flipped, a corresponding flipping operation is applied to the MLP output. Full implementation details of the above procedures are provided in the Methods.
Whilst we focus on encoding the specific spin-flip invariance required by the ARN, the core idea of encoding general symmetries into a NNQS by mapping related configurations to a single canonical form is explored in detail in~\cite{choo_symmetries}.

The spin-flip symmetry is only meaningful for the molecules being optimised with $M_S{=}0$, as this is required for both $\vb{x}_{k}$ and $F(\vb{x}_{k})$ to be physically valid configurations. With our choice of $M_S{=}S$, this constraint could only be applied to closed shell molecules with $S{=}0$. Additionally, we would not have been able to apply the spin-flip symmetry if our basis of spin-orbitals did not restrict their spatial component to be independent of the electron spin (as generated by so called unrestricted Hartree-Fock methods).

\begin{table*}[t]
\centering
\resizebox{\textwidth}{!}{%
\begin{tabular}{ccccc cccccc ccc}

\toprule

\multicolumn{5}{c}{Molecular information} &
\multicolumn{6}{c}{Method comparison} &
\multicolumn{3}{c}{NAQS variants} \\

\cmidrule[0.75pt](lr){1-5} \cmidrule[0.75pt](lr){6-11} \cmidrule[0.75pt](lr){12-14} 

Molecule & $N_{\mathrm{O}}$ & $N_{\mathrm{e}}$ & $S$ & Valid $\vb{x}_{k}$ & HF & CISD & CCSD & CCSD(T) & NAQS & FCI & Standard & No mask & No\,spin\,sym.
\\

\cmidrule(lr){1-5} \cmidrule(lr){6-11} \cmidrule(lr){12-14} 

\ce{H2} &
	 4 & 2 & 0 & \num{4} &
	 
	 -0.9109 & 
	 \underline{\bf{-0.9981}} & 
	 \underline{\bf{-0.9981}} & 
	 \underline{\bf{-0.9981}} & 
	 \underline{\bf{-0.9981}} & 
	 -0.9981 & 
	 
	 \underline{\bf{-0.9981}} &
	 \underline{\bf{-0.9981}} & 
	 \underline{\bf{-0.9981}}
	 \\
	 
\ce{F2} &
	20 & 18 & 0 & \num{100} &
	
	 -195.6380 & 
	 \underline{\bf{-195.6611}} & 
	 \underline{\bf{-195.6611}} & 
	 \underline{\bf{-195.6611}} &
	 \underline{\bf{-195.6611}} & 
	 -195.6611 &
	 
	 \underline{\bf{-195.6611}} &
	 \underline{\bf{-195.6611}} & 
	 \underline{\bf{-195.6611}} 
	 \\
	 
\ce{HCl} & 
	20 & 18 & 0 & \num{100} &
	
	 -455.1360 & 
	 \underline{\bf{-455.1562}} &
	 \underline{\bf{-455.1562}} & 
	 \underline{\bf{-455.1562}} &
	 \underline{\bf{-455.1562}} &
	 -455.1562 & 
	 
	 \underline{\bf{-455.1562}} &
	 \underline{\bf{-455.1562}} & 
	 \underline{\bf{-455.1562}} 
	 \\
	 
\ce{LiH} &
	 12 & 4 & 0 & \num{225} &
	 
	 -7.7674 & 
	 \underline{\bf{-7.7845}} & 
	 \underline{\bf{-7.7845}} & 
	 \underline{\bf{-7.7845}} &
	 \underline{\bf{-7.7845}} &
	 -7.7845 & 
	 
	 \underline{\bf{-7.7845}}	& 	 	 
	 \underline{\bf{-7.7845}} & 
	 \underline{\bf{-7.7845}} 
	 \\
	 
\ce{H2O} & 
	 14 & 10 & 0 & \num{441} &
	 
	 -74.964 & 
	 -75.0148 & 
	 -75.0151 & 
	 \underline{\bf{-75.0155}} &
	 \underline{\bf{-75.0155}} &
	 -75.0155 & 
	 
	 \underline{\bf{-75.0155}} &
	 \underline{\bf{-75.0155}} & 
	 \underline{\bf{-75.0155}} 
	 \\
	 
\ce{CH2} & 
	 14 & 8 & 1 & \num{735} &
	 
	 -37.4846 &
	 \underline{\bf{-37.5044}} & 
	 \underline{\bf{-37.5044}} &
	 \underline{\bf{-37.5044}} &
	 \underline{\bf{-37.5044}} &
	 -37.5044 &
	 
	 --- &
	 \underline{\bf{-37.5044}} & 
	 \underline{\bf{-37.5044}} 
	 \\
	 
\ce{O2} &
	20 & 16 & 1 & \num{1200} &
	
	 -147.6319 & 
	 \underline{\bf{-147.7502}} & 
	 -147.7477 & 
	 -147.7485 &
	 -147.7500 &
	 -147.7502 & 
	 
	 --- &	 
	 -147.7496 & 
	 -147.7500 
	 \\
	 
\ce{BeH2} &
	14 & 6 & 0 & \num{1225} &
	
	 -14.4432 & 
	 -14.4725 & 
	 -14.4727 & 
	 \underline{\bf{-14.4729}} &
	 \underline{\bf{-14.4729}} & 
	 -14.4729 & 
	 
	 \underline{\bf{-14.4729}} & 
	 \underline{\bf{-14.4729}} & 
	 \underline{\bf{-14.4729}} 
	 \\
	 
\ce{H2S} & 
	22 & 18 & 0 & \num{3025} &
	
	 -394.3114 & 
	 -394.3539 &
	 \underline{\bf{-394.3546}} &
	 \underline{\bf{-394.3546}} &
	 \underline{\bf{-394.3546}} & 
	 -394.3546 & 
	 
	 \underline{\bf{-394.3546}} & 	 
	 \underline{\bf{-394.3546}} & 
	 \underline{\bf{-394.3546}} 
	 \\
	 
\ce{NH3} &
	16 & 10 & 0 & \num{3136} &
	
	 -55.4548 & 
	 -55.5195 & 
	 -55.5209 & 
	 -55.5210 &
	 \underline{\bf{-55.5211}} &
	 -55.5211 & 
	 
	 \underline{\bf{-55.5211}} &
	 \underline{\bf{-55.5211}} & 
	 \underline{\bf{-55.5211}} 
	 \\
	 
\ce{N2} &
	20 & 14 & 0 & \num{14400} &
	
	 -107.4990 & 
	 -107.6471 & 
	 -107.6561 & 
	 -107.6579 &
	 \bf{-107.6595} &
	 -107.6602 & 
	 
	 \bf{-107.6595} &
	 -107.6588 & 
	 -107.6511 
	 \\
	 
\ce{CH4} & 
	20 & 10 & 0 & \num{15876} &
	
	 -39.7266 & 
	 -39.8035 &
	 -39.8060 & 
	 \bf{-39.8062} &
	 \bf{-39.8062} &
	 -39.8063 & 
	 
	 \bf{-39.8062} &	 
	 -39.8061 & 
	 -39.8061 
	 \\	
		 
\ce{C2} &
	20 & 12 & 0 & \num{44100} &
	
	 -74.4209 & 
	 -74.6371 & 
	 -74.6745 & 
	 -74.6876 &
	 \bf{-74.6899} &
	 -74.6908 & 
	 
	 \bf{-74.6899} &	 
	 -74.6562 & 
	 -74.6898 
	 \\
	 
\ce{LiF} & 
	20 & 12 & 0 & \num{44100} &
	
	 -105.1137 & 
	 -105.1607 &
	 -105.1592 & 
	 \emph{-105.1663} &
	 \underline{\bf{-105.1662}} &
	 -105.1662 & 
	 
	 -105.1660 &
	 \underline{\bf{-105.1662}} & 
	 -105.1660 
	 \\
	 
\ce{PH3} & 
	24 & 18 & 0 & \num{48400} &
	
	 -338.6341 & 
	 -338.6963 &
	 -338.6982 & 
	 \underline{\bf{-338.6984}} &
	 \underline{\bf{-338.6984}} &
	 -338.6984 & 
	 
	 \underline{\bf{-338.6984}} & 
	 \underline{\bf{-338.6984}} & 
	 \underline{\bf{-338.6984}} 
	 \\
	 	 
\ce{LiCl} & 
	28 & 20 & 0 & \num{1002001} &
	
	 -460.8273 & 
	 -460.8482 &
	 -460.8476 & 
	 \emph{-460.8500} &
	 \underline{\bf{-460.8496}} &
	 -460.8496 & 
	 
	 \underline{\bf{-460.8496}} &
	 \underline{\bf{-460.8496}} & 
	 \underline{\bf{-460.8496}} 
	 \\
	 
\ce{Li2O} & 
	30 & 14 & 0 & \num{41409225} &
	
	 -87.7956 & 
	 -87.8837 &
	 -87.8855 & 
	 \emph{-87.8931} &
	 \bf{-87.8909} &
	 -87.8927 & 
	 
	 -87.8878 &
	 \bf{-87.8909} & 
	 -87.8867 
	 \\
	 
\bottomrule

\end{tabular}
} 
\caption{
Obtained molecular ground-state energies (in \si{Ha}) using various canonical methods and a variationally optimised NAQS (``Standard'' following the procedure from the main text, along with associated ablations).  Also provided is information for each molecule, including the number of physically valid determinants (Valid $\vb{x}_{k}$) remaining after reducing the optimisation space obtained using the Jordan-Wigner transform to conserve electron number, $N_{\mathrm{e}}$, and total spin, $S$.
All stochastic methods (NAQS and associated ablations) report the best result obtained across five optimisation attempts, with the exception of \ce{Li2O} for which only a single attempt was used due to the increased computational cost.
The best non-FCI results are displayed in bold (with those matching FCI additionally underlined), excluding methods that produce unphysical energies below the FCI limit, which are displayed in italic.
}
\label{tab:performance_comps_app}
\end{table*}

\section{Results}

We now investigate both the performance of our network across a range of molecules and the importance of particular aspects of our architecture.  To avoid ``cherry-picking'', all results presented in this work use a single set of network hyperparameters and identical training procedures.

\subsection{Obtained molecular ground states}

The full experimental results obtained are presented in \cref{tab:performance_comps_app} from which we can see that our neural autogressive quantum state (NAQS) exhibits consistently strong performance on all molecules considered.
As baseline approximate QC methods, we consider the Hartree-Fock energy (HF), CI with single and double excitations (CISD), and CC with up to both double (CCSD) and perturbative triple excitations (CCSD(T)).

\begin{figure}[!b]
\centering
\includegraphics[width=\columnwidth]{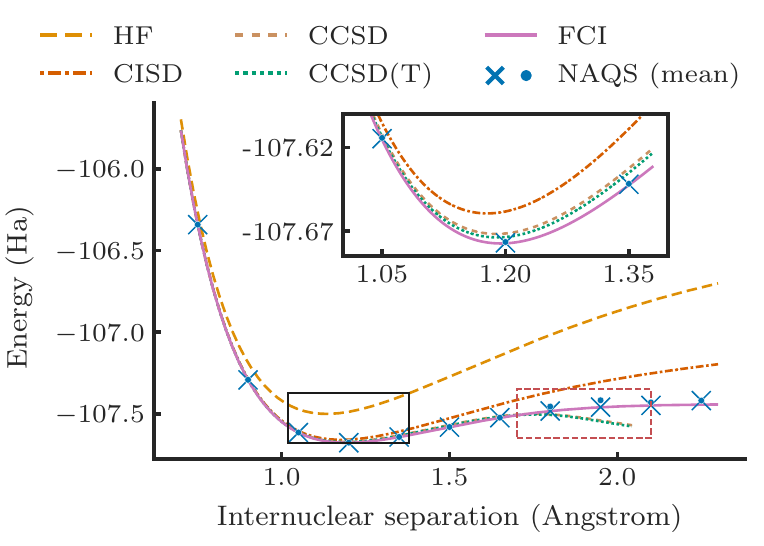}
\caption{Comparison of the energy obtained using a neural autoregessive quantum state (NAQS) to traditional QC approaches for the diatomic nitrogen molecule, as a function of the nuclear separation.  The NAQS outperforms all other approximation techniques, almost exactly matching the ground-truth FCI solution in all cases, including near equilibrium (inset and solid black box) and even away from the equilibrium geometry where other methods --- including the otherwise most accurate coupled-cluster approach --- fail due to the presence of strong quantum correlations (highlighted with dashed-red box).  The best (crosses) and mean (dots) NAQS results across 5 seeds are presented at each distance.}
\label{fig:N2energySurface}
\end{figure}

The obtained NAQS energies approach or match the ground-truth FCI result on all molecules with up to 20 electrons and 28 spin-orbitals, even when the most sophisticated baselines do not.  Indeed, our NAQS still outperforms CCSD and CISD on \ce{Li2O} which has ${>}\SI{40}{M}$ physically valid basis determinants under the Jordan-Wigner transformation.  An apparent exception is a lower energy for \ce{Li2O} provided by CCSD(T); however, this result is also below the FCI limit (and is thus unphysical).  In contrast, the ARN ansatz guarantees physical validity of its output states.
This is further illustrated in \cref{fig:N2energySurface} where FCI-like accuracy is obtained for the energy surface of \ce{N2}, even in regimes where the CC baselines fail.  Ultimately, this highlights the high degree of correlations and entanglement the NAQS is capable of learning.

It is illuminating to compare these results to the best (and, indeed, only) previous work applying neural network quantum states to quantum chemistry in the second quantisation -- the RBM-based ansatz of Choo~\emph{et~al.}~\cite{choo20}.  The largest system on which it surpasses coupled-cluster methods is \ce{C2} (\SI{44.1}{k} determinants), with the sampling-limited performance on larger molecules beating CISD on a system with ${\sim}\SI{1.6}{M}$ determinants.  Our ansatz scales to systems more than an order of magnitude larger without a loss of performance, and outperforms the results of Ref.~\cite{choo20} on every molecule larger than \ce{H2}, which is trivially solved by most methods (numerical results are provided in the Supplementary Information). 

Interestingly, we observe ``under-sampling'' of the optimisation space for large molecules.  Specifically, with (without) masking unphysical determinants, for \ce{LiCl} and \ce{Li2O} we typically only sampled ${\sim}\SI{28}{\percent}$ (\SI{20}{\percent}) and ${\sim}\SI{5}{\percent}$ (\SI{1}{\percent}) of the physically viable determinants at least once during training, respectively, and still learnt the important configurations required for accurate wavefunctions. This suggests that the neural network is able to generalise to  unseen configurations, and therefore offers a promising approach to efficiently exploring and isolating the important components of the entire optimisation space.

We observed that after an initial period of exploration, the ARN typically isolates a relatively small number of determinants that dominate the NAQS. 
The remainder of the optimization run is spent fine-tuning the amplitudes of these determinants.
For example, even without applying masking to restrict the ARN to only physically viable determinants, the final step of \ce{Li2O} generated $10^{12}$ samples distributed across only \num{13761} unique determinants. If we take the subspace spanned by these determinants and simply diagonalise the Hamiltonian, we obtain an energy of \SI{-87.8911}{Ha}, which is even lower than that found by the ARN (\SI{-87.8909}{Ha}).
This suggests a hybrid approach, where canonical numerical methods are applied to the subspace of configurations found by the network to achieve even lower energies.
Further exploration of these ideas is left for future work.


\subsection{Ablation studies}

To examine the impact of encoding physical priors into the NAQS, \cref{tab:performance_comps_app} additionally includes results for two ablations.  A NAQS with ``No mask'' can assign non-zero probability to, and thus sample, any of the $2^{N_{\mathrm{O}}}$ possible configurations regardless of whether they are physically viable.  ``No spin sym.'' indicates that the spin-flip invariance (\ref{eq:spinFlipSym}) is removed from the wavefunction.  An exception is open-shell molecules (\ce{CH2} and \ce{O2}), to which spin-flip invariance could not be applied; however, these molecules showed FCI-level performance even without this constraint. 

\begin{figure}[!t]
\centering
\includegraphics[width=\columnwidth]{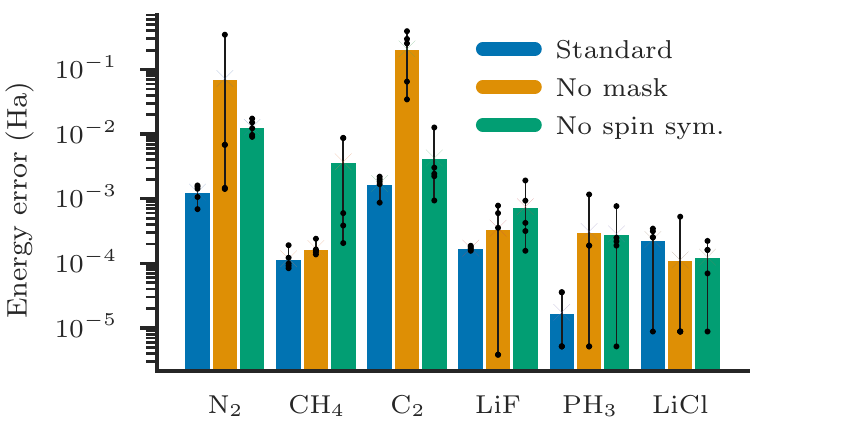}
\caption{The variational energies obtained over the course of optimisation for the NAQS model described in the text (``Standard'') and the associated ablations of both restricting the optimisation space to physically viable determinants (``No mask'') and the spin-flip symmetries of the final wavefunction (``No spin sym.'').  Each bar denotes the average performance across the 5 training seeds, with the dots denoting the energies obtained by individual runs and the vertical black line indicating the span from the best to worst of these.  Where fewer than 5 dots are visible, this is due the same energy being found multiple times.  Energies are plotted in terms of their ``error'', the distance from the ground-truth FCI energy, on a logarithmic scale.}
\label{fig:meanPerformance}
\end{figure}

Whilst we see that removing either of these physical priors can result in a slight drop in performance, the effect appears subtle.  However, when we consider the performance across multiple initialisations a clearer distinction is evident.  
\Cref{fig:meanPerformance} shows the energies obtained for every individual run, and the corresponding overall average performance, on the six largest molecules for which we performed multiple optimisations.  We can see that the average performance is significantly improved when our physical knowledge is encoded into the system.  Moreover, in the few cases that the best result is obtained by an ablation (e.g.\@ ``No mask'' on \ce{LiF}) it is clear that this is the result of one fortuitous seed, rather than a systematic effect.

In general, the effect of these ablations only becomes visible on larger systems, such as those shown in \cref{fig:meanPerformance}.  This seems reasonable, as with the larger dimension of the Hilbert space the set of possible configuration strings grows exponentially, hence it becomes more critical to reduce its size with the known \emph{a priori} constraints.

\section{Discussion}

This work presents an ARN ansatz for \emph{ab initio} QC that approaches FCI-quality solutions on systems with up to 30 spin orbitals, even in situations where canonical approximation methods fail.  This success is attributable to both the ability to embed physical priors into the network -- specifically, spin-flip invariance, and conservation of electron number and total spin --  and an efficient sampling procedure that scales with the number of unique configurations sampled, rather than the total batch size.  These advances ultimately allow our approach to scale to systems far beyond what was previously possible in ML-based approaches to second quantised QC~\cite{choo20}.

There are two prominent factors in the overall optimisation speed (and thus scalability) of our ARN: (\romannumeral 1) sampling configurations from the wavefunction and (\romannumeral 2) evaluating the local energies.  Analysis of the optimisation process (detailed in the Supplementary Information) shows that (\romannumeral 1) is determined by the cost of a network forward pass, which scales as $\mathcal{O}(MLh^3)$ for $M{=}N_{\mathrm{O}}/2$ orbitals, and $L$ layers of dimension $h$.  When few unique determinants are sampled (e.g.\@ in the case of small molecules) this is the rate-determining step.  However, given a batch of $N_{\mathrm{unq}}$ unique samples, calculating the local energies, (\romannumeral 2), as per \cref{eq:Eloc} scales as $\mathcal{O}(N_{\mathrm{O}}^{4}N_{\mathrm{unq}})$ for second-quantised QC Hamiltonians~\cite{mcardle20}.  As a result, in practice, this calculation becomes the dominant computational expense for large molecules.  Therefore, whilst this work addresses the requirement for an efficient and expressive ansatz, the remaining barrier to even larger molecules is the computational cost of VMC calculations separate from neural network itself.

It is also important to note that the fundamental accuracy of a second-quantized approach is limited by the choice of basis set, with the challenges associated with scaling to larger basis sets equivalent to those of scaling to larger molecules discussed above. 
An alternative approach to overcome this limitation is to consider the electronic structure problem in first quantization and use neural networks to encode real-space wavefunctions with flexible basis sets, thus encoding the ground state in considerably fewer determinants.  Two eminent examples are FermiNet~\cite{pfau19} and PauliNet~\cite{hermann19}, both of which report accuracies unconstrained by a fixed basis set, albeit at the expense of increased computational complexity.
Whilst these approaches both use MCMC sampling, their compatibility with our second-quantized picture is not clear, however this comparison remains an interesting prospect for future research.

\small

\section*{Acknowledgements}

The authors are grateful to G.~Carleo for his insights regarding RBMs, and to M.~Sapova for her assistance with quantum chemical calculations.  A.L.'s research is partially supported by Russian Science Foundation (19-71-10092).

\section*{Author contributions}

T.B.\ conceived the research, wrote the code, performed the experiments and co-wrote the paper.  A.M.\ assisted on theoretical analysis of system and in preparing the manuscript.  A.L.\ oversaw the entire project, helped interpreting the results and co-wrote the paper.

\section*{Competing interests}

The authors declare no competing interests.

\section*{Data availability}

No datasets were generated or analysed during the current study.

\section*{Code availability}

Source code for this work, including experimental scripts and molecular data required to reproduce the reported results, can be found at \url{https://github.com/tomdbar/naqs-for-quantum-chemistry}.

\section*{Methods}

\subsection*{Mapping fermionic Hamiltonians to the qubits}

We consider the second quantised form of the molecular electronic Hamiltonian~\cite{mcardle20},
\begin{equation}
	\mathcal{H} = \sum_{p,q} h_{pq} \hat{a}^{\dag}_p\hat{a}_q + \frac{1}{2}\sum_{p,q,r,s} h_{pqrs} \hat{a}^{\dag}_p\hat{a}^{\dag}_q\hat{a}_r\hat{a}_s,
	\label{eq:Ham2ndQuan}
\end{equation}
where $\hat{a}^{\dag}_i$ ($\hat{a}_i$) is fermionic electron creation (annihilation) operator for the $i$-th single-electron orbital and $h_{pq}$, $h_{pqrs}$ are the one- and two-body integrals, respectively.  We use the Jordan-Wigner encoding~\cite{wigner28} to map from operators acting on indistinguishable fermions to operators acting on distinguishable qubits.  This is given by
\begin{equation}
	\hat{a}^{\dag}_j \rightarrow \left( \prod_{i=0}^{j-1} \sigma^{z}_{i} \right) \sigma^{+}_{j},
	\quad
	\hat{a}_j \rightarrow \left( \prod_{i=0}^{j-1} \sigma^{z}_{i} \right) \sigma^{-}_{j},
	\label{eq:jwMapping}
\end{equation}
which stores the occupation state of the $j$-th fermionic mode locally in the $j$-th qubit --- using the raising/lowering operator $\sigma^{\pm}_{j}= \sigma^{x}_{j} \pm \mathrm{i} \sigma^{y}_{j}$ --- whilst the parity is stored non-locally.
Whilst other mappings that differently store the occupation and parity of fermionic modes could be used --- such as the parity~\cite{seeley12} or Bravyi-Kitaev~\cite{bravyi02} encodings --- these would each require bespoke consideration of how to encode physical priors into our system.
Regardless of the choice of encoding, the qubit Hamiltonian still takes the form of \cref{eq:HamSpin} from the main text.

\subsection*{Network architecture}

All results presented in this work use a single set of network hyperparameters.  Each subnetwork, $\ln \abs*{\psi_{i}(v_{k}^{i} \vert v_{k}^{1}, \dots, v_k^{i-1} )}$, uses an MLP with a single hidden layer of 64 neurons with the single phase network, $\phi(\cdot)$, using two hidden layers of 512 neurons.
Both networks use ReLU activations~\cite{glorot11} on all hidden layers, with no activation on the input or output layers.  We emphasise that this architecture is massively over-parameterised for most, if not all, of the molecules we consider.  However, the forward and backward pass of the network are not our computational bottleneck (as discussed, it was the calculation of matrix-elements of the Hamiltonian) and therefore optimising the network size was not considered.

The autoregressive wavefunction decomposition defined in \cref{eq:arWavefunction} does leave a degree of freedom in the form of the order of our conditional wavefunctions.  In this work, we consider orbitals in order of decreasing energy, i.e.\@ $\ln \psi_{1}(\cdot)$ and $\ln \psi_{M}(\cdot)$ correspond to the highest and lowest energy orbitals, respectively.  Optimal ordering of the spin-orbitals in a NAQS can be an interesting subject for future research. 

To ensure the conditional wavefunction have spin-flip symmetry as defined in \cref{eq:spinFlipSym} of the main text, we first split the input to the $i$-th subnetwork into its spin-up and spin-down orbitals: $( x_{k}^{1\uparrow}, \dots, x_{k}^{(i-1)\uparrow} )$ and $( x_{k}^{1\downarrow}, \dots, x_{k}^{(i-1)\downarrow} )$.  Recalling that $x_{k} \in \{0,1\}$, we can consider each of these partially sampled configurations as a binary-encoded integer. 

The input to each sub-network is therefore invariant to global spin flips, so we next have to reconstruct the desired symmetries at the output.  Concretely, the MLP outputs 5 values
\begin{equation}
	\text{MLP}_i(R(v_{k}^{1} \dots v^{k}_{i-1})) = [ z_1, z_2, z_3, z_4, z_5 ], \quad z_j \in \mathbb{R},
\end{equation}
which we convert to four unnormalised log-probabilty amplitudes for the value of the next pair of qubits.  If we have a spin-symmetric input ($F(v_{k}^{1} \dots v^{k}_{i-1})=(v_{k}^{1} \dots v^{k}_{i-1})$), then the outputs are simply
\begin{align*}
	\ln \abs*{\tilde\psi_i(0,0)} &= z_1 \\
	\ln \abs*{\tilde\psi_i(0,1)} &= z_2 \\
	\ln \abs*{\tilde\psi_i(1,0)} &= z_2 \\
	\ln \abs*{\tilde\psi_i(1,1)} &= z_3
\end{align*}
(where the tilde indicates that the amplitudes are not normalized). However, if the input is not spin-symmetric ($F(v_{k}^{1} \dots v^{k}_{i-1}) \neq (v_{k}^{1} \dots v^{k}_{i-1})$), then the outputs are
\begin{align*}
	\ln \abs*{\tilde\psi_i(0,0)} &= z_1 \\
	\ln \abs*{\tilde\psi_i(0,1)} &= 
		\begin{cases}
			(z_2 {+} z_4)/2 &\text{if } n(v_{k}^{1} \dots v_{k}^{i-1})<n(F(v_{k}^{1} \dots v_{k}^{i-1})) \\
			(z_2 {+} z_5)/2 &\text{if } n(v_{k}^{1} \dots v_{k}^{i-1})>n(F(v_{k}^{1} \dots v_{k}^{i-1})) \\
		\end{cases} \\
	\ln \abs*{\tilde\psi_i(1,0)} &= 
		\begin{cases}
			(z_2 {+} z_5)/2 &\text{if } n(v_{k}^{1} \dots v_{k}^{i-1})<n(F(v_{k}^{1} \dots v_{k}^{i-1})) \\
			(z_2 {+} z_4)/2 &\text{if } n(v_{k}^{1} \dots v_{k}^{i-1})>n(F(v_{k}^{1} \dots v_{k}^{i-1})) \\
		\end{cases} \\
	\ln \abs*{\tilde\psi_i(1,1)} &= z_3.
\end{align*}
These definitions satisfy the symmetries discussed in the main text.  Namely, a spin-symmetric input ($F(v_{k}^{1} \dots v^{k}_{i-1})=(v_{k}^{1} \dots v^{k}_{i-1})$) should have spin-symmetric outputs, and that a pair of spin-symmetric inputs should have correspondingly symmetric outputs.  Of course, there are many choices for how to combine or permute outputs of an MLP to achieve these effects, with the above being what we found to work best in practice.   

\subsection*{Training details}

The network can be trained by following the gradients defined in \cref{eq:VMCgrad}.  However, the variance of the gradients can be reduced, without introducing any bias into the sample approximation, by subtracting a baseline~\cite{allah20},
\begin{equation}
	\nabla_\theta E =
	2\Re \left( \mathbb{E}_{p} \big[
	(E_{\mathrm{loc}} - \mathbb{E}_{p} [ E_{\mathrm{loc}} ])\nabla_\theta \ln(\psi^{\ast})
	\big] \right).
	\label{eq:VMCgrad_baselined}
\end{equation}
This improves training stability and is a standard technique deployed in machine learning to improve gradient estimation~\cite{sutton18, mohamed19}.

Training follows the standard VMC approach described in the main text, with all parameters simultaneously optimised using the Adam algorithm~\cite{kingma14} with an initial learning rate of $5\times10^{-3}$, dropped to $5\times10^{-4}$ halfway through training, and decay rates for the first- and second-moment estimates of $\beta_1=0.9$ and $\beta_2=0.99$, respectively.  The number of samples ($N$ in \cref{alg:sampling}) used to estimate the stochastic gradient steps, \cref{eq:VMCgrad_baselined}, was automatically adjusted to ensure the number of unique physically valid samples remained between \SI{10}{k} and \SI{100}{k} for as long as possible.  Specifically, the initial batch size of $10^6$ was increased/decreased by an order of magnitude whenever the additional/fewer unique samples were required, with an upper batch size limit of $10^{12}$.

The random initialisation of the network parameters (which was not modified from the default behaviour of PyTorch and can be found in the supporting code) does impact final performance, therefore to avoid ``cherry-picking'', all reported results are the best energy obtained across exactly 5 seeds, each trained for \SI{10}{k} steps, with the exception of \ce{Li2O} which uses only a single seed due to the increased computational cost.

\subsection*{Molecular geometries and calculations}

The molecular geometries used are the geometries in the STO-3G basis returned from the PubChem~\cite{kim16} database by OpenFermion~\cite{mcclean20}.  OpenFermion was also used to generate qubit Hamiltonians of the form of \cref{eq:HamSpin}, with the backend calculations and baseline QC methods -- HF, CI, CCSD, CCSD(T) --- implemented using Psi4~\cite{smith20}.  Exact geometries and scripts to reproduce all of these steps can be found in the supporting code.

\renewcommand*{\bibfont}{\small}
\bibliographystyle{apsrev4-1}
\bibliography{References}

\clearpage

\onecolumngrid

\section*{Supplementaty Information}

\setcounter{table}{0}
\setcounter{figure}{0}
\setcounter{section}{0}

\section{Further details on results from main text}

\subsubsection*{Full comparison of NAQS and RBMs}

To directly compare the performance of our NAQS agains the RBM of Choo~\etal{}~\cite{choo20}, we applied our NAQS to exactly the molecular geometries presented in their work.  These results are summarised in \cref{tab:performance_comps_mean_carleo}, where we see that our NAQS outperforms \cite{choo20} on every molecule larger that \ce{H2} (which is trivially solved by both).

\begin{table*}[h]
\centering
\begin{tabular}{cccccccc}

\toprule

&
\multicolumn{4}{c}{Classical methods} &
\multicolumn{2}{c}{NNQS} \\

\cmidrule[0.75pt](lr){2-5} \cmidrule[0.75pt](lr){6-7} 

Molecule & HF & CISD & CCSD & CCSD(T) & RBM & NAQS & FCI
\\

\cmidrule(lr){1-8}

\ce{H2} &
	 -1.1170 &
	 \underline{\bf{-1.1373}} &
	 \underline{\bf{-1.1373}} &
	 \underline{\bf{-1.1373}} &
	 \underline{\bf{-1.1373}} &
	 \underline{\bf{-1.1373}} &
	 -1.1373
	 \\
	 
\ce{LiH} &
	 -7.8631 &
	 -7.8827 &
	 \underline{\bf{-7.8828}} &
	 \underline{\bf{-7.8828}} &
	 -7.8826 &
	 \underline{\bf{-7.8828}} &
	 -7.8828
	 \\
	 
\ce{H2O} &
	 -74.9625 &
	 -75.0221 &
	 -75.0231 &
	 -75.0232 &
	 -75.0232 &
	 \underline{\bf{-75.0233}} &
	 -75.0233
	 \\
	 
\ce{NH3} &
	 -55.4513 &
	 -55.5258 &
	 -55.5280 &
	 -55.5281 &
	 -55.5277 &
	 \underline{\bf{-55.5282}} &
	 -55.5282
	 \\
	 
\ce{N2} &
	 -107.4912 &
	 -107.6591 &
	 -107.6717 &
	 -107.6738  &
	 -107.6767 &
	 \underline{\bf{-107.6774}} &
	 -107.6774
	 \\
	 
\ce{C2} &
	 -74.4209 &
	 -74.6371 &
	 -74.6745 &
	 -74.6876 &
	 -74.6892 &
	 \bf{-74.6895} &
	 -74.6908
	 \\
	 	 
\bottomrule

\end{tabular}
\caption{
Comparison of the performance a NAQS and an RBM.  NAQS results were obtained by following the exact same procedure as described for the results in the main text.  RBM results are as reported by Choo~\emph{et al.} in reference [25] of the main text.  The best results are displayed in bold (with those matching FCI additionally underlined).  Note that the molecular geometries match those from [25], which differ from those presented in Table 1 of the main text, hence the different FCI energies for the same species of molecule.}
\label{tab:performance_comps_mean_carleo}
\end{table*}

\subsubsection*{Average performance of ablations}

The we performed an ablation study on the physical priors embedded into the NAQS, with numerical results provided in Table I of the main text and the variance of these models over multiple seeds summarised in Figure 4 of the main text.  For completeness, we here provide the average performance across multiple seeds in \cref{tab:performance_comps_mean}.

\begin{table*}[h]
\centering
\begin{tabular}{cccccc ccc}

\toprule

\multicolumn{6}{c}{Molecular information} &
\multicolumn{3}{c}{NAQS variants} \\

\cmidrule[0.75pt](lr){1-6} \cmidrule[0.75pt](lr){7-9} 

Molecule & $N_{\mathrm{O}}$ & $N_{\mathrm{e}}$ & $S$ & Valid $\vb{x}_k$ & FCI & Standard & No mask & 
No\,spin\,sym.
\\

\cmidrule(lr){1-6} \cmidrule(lr){7-9}

\ce{H2} &
	 4 & 2 & 0 & \num{4} &
	 
	 -0.9981 &
	 
	 \underline{\bf{-0.9981}} & 
	 \underline{\bf{-0.9981}} & 
	 \underline{\bf{-0.9981}} 
	 \\
	 
\ce{F2} &
	20 & 18 & 0 & \num{100} &

	 -195.6611 &
	 
	 \bf{-195.6605(12)} & 
	 -195.6579(48) & 
	 -195.660(1) 
	 \\
	 
\ce{HCl} & 
	20 & 18 & 0 & \num{100} &
	
	 -455.1562 &
	 
	 \underline{\bf{-455.1562}} & 
	 -455.1561(1) & 
	 \underline{\bf{-455.1562}} 
	 \\
	 
\ce{LiH} &
	 12 & 4 & 0 & \num{225} &

	 -7.7845 &
	 	 
	 \underline{\bf{-7.7845}} & 
	 \underline{\bf{-7.7845}} & 
	 -7.7845(7) 
	 \\
	 
\ce{H2O} & 
	 14 & 10 & 0 & \num{441} &
	 
	 -75.0155 &
	 
	 \underline{\bf{-75.0155}} & 
	 \underline{\bf{-75.0155}} & 
	 \underline{\bf{-75.0155}} 
	 \\
	 
\ce{CH2} & 
	 14 & 8 & 1 & \num{735} &

	 -37.5044 &
	 
	 --- & 
	 -37.33(23) & 
	 \bf{-37.5041(4)} 
	 \\
	 
\ce{O2} &
	20 & 16 & 1 & \num{1200} &

	 -147.7502 &
	 	 
	 --- & 
	 -147.70(92) & 
	 -147.747(3) 
	 \\
	 
\ce{BeH2} &
	14 & 6 & 0 & \num{1225} &
	
	 -14.4729 &
	 
	 \underline{\bf{-14.4729}} & 
	 \underline{\bf{-14.4729}} & 
	 -14.471(3) 
	 \\
	 
\ce{H2S} & 
	22 & 18 & 0 & \num{3025} &

	 -394.3546 &
	 	
	 \underline{\bf{-394.3546}} & 
	 -394.35(13) & 
	 \underline{\bf{-394.3546}} 
	 \\
	 
\ce{NH3} &
	16 & 10 & 0 & \num{3136} &

	 -55.5211 &
	 
	 \underline{\bf{-55.5211}} & 
	 \underline{\bf{-55.5211}} & 
	 \underline{\bf{-55.5211(1)}} 
	 \\
	 
\ce{N2} &
	20 & 14 & 0 & \num{14400} &

	 -107.6602 &
	 
	 \bf{-107.6589(4)} & 
	 -107.65(15) & 
	 -107.647(4) 
	 \\
	 
\ce{CH4} & 
	20 & 10 & 0 & \num{15876} &

	 -39.8063 &
	 	
	 \bf{-39.8061} & 
	 \bf{-39.8061} & 
	 -39.803(5) 
	 \\	
		 
\ce{C2} &
	20 & 12 & 0 & \num{44100} &

	 -74.6908 &

	 \bf{-74.6891(5)} & 
	 -74.48(15) & 
	 -74.687(5) 
	 
	 \\
	 
\ce{LiF} & 
	20 & 12 & 0 & \num{44100} &

	 -105.1662 &
	 
	 \bf{-105.1660} & 
	 -105.1658(3) & 
	 -105.1654(7) 
	 \\
	 
\ce{PH3} & 
	24 & 18 & 0 & \num{48400} &

	 -338.6984 &
	 	
	 \underline{\bf{-338.6984}} & 
	 -338.6981(5) & 
	 -338.6981(3) 
	 \\
	 	 
\ce{LiCl} & 
	28 & 20 & 0 & \num{1002001} &
	
	 -460.8496 & 
	 
	 -460.8494(1) & 
	 -460.8495(2) & 
	 \bf{-460.8495(1)} 
	 \\
	 
\bottomrule

\end{tabular}
\caption{
Average molecular ground-state energies (in \si{Ha}) of different NAQS configurations as described in the main text.  Averages are across 5 seeds with brackets denoting the standard deviation across all seeds on the final digits.  Where no uncertainty is denoted, the variation is below the level of precision reported in the table.  Notably large uncertainties, such as `No mask' for \ce{O2}, are the result of rare failure seeds, where a single run of NAQS fails to converge to a meaningful energy.
The `missing' entries for \ce{CH2} and \ce{O2} indicate that we do not enforce spin-symmetry to these systems as discussed in the main text. Also provided is information for each molecule, including the number of physically valid determinants (`Valid $\vb{x}_{k}$') remaining after reducing the optimisation space obtained using the Jordan-Wigner transform to conserve electron number, $N_{\mathrm{e}}$, and total spin, $S$.
The best results are displayed in bold (with those matching FCI additionally underlined).
}
\label{tab:performance_comps_mean}
\end{table*}

\sisetup{
  output-exponent-marker = \text{e},
  table-format=+1.4e+2,
  exponent-product={},
  retain-explicit-plus
}

\section{Analysis of wall-clock timings}

Here, we detail the timings of each step of our optimisation process and, in particular, the practical scaling of these costs with system size.  Unless otherwise stated, all calculations were performed on GeForce RTX 2080 Ti (11\,GB) and 32 Intel(R) Xeon(R) CPU E5-2620 v4 processors (shared with other users).

\subsubsection*{Pre-calculating coupling terms.}

Our starting point is the Jordan-Wigner encoded Hamiltonian~\cite{wigner28},
\begin{equation}
	\Ham\sub{Q} = \sum_{j=1}^{K} h_j \prod_{i=1}^{\No} \sig{\nu_{j,i}}{i},
	\label{eq:JWham}
\end{equation}
where $\sig{\nu_{j,i}}{i}$ is a Pauli operator acting on the $i$-th qubit ($\nu_{j,i} \in \{ \mathrm{I}, x, y, z \}$), as this is readily obtained using Psi4~\cite{smith20}.  To convert this Hamiltonian to a matrix form, where each element couples a pair of Slater determinants, we calculate each element using the procedure described in \cite{choo20}.
The total time to calculate all non-zero elements of the Hamiltonian coupling matrix depends on both the number of `Pauli strings', $K$, in the encoded Hamiltonian (\ref{eq:JWham}) and the number of physically valid (and thus possibly coupled) configurations.  These metrics are summarised, along with the total conversion time for the entire Hamiltonian, in \cref{tab:coupling_calcs}.  We see that the cost of this calculation scales with system size, taking the order of minutes for \ce{LiCl}.  Moreover, in practice we never converted the full Hamiltonian for \ce{Li2O} due to the excessive cost, so instead these couplings were computed on demand and cached during the optimisation. 

\begin{table*}[h]
\centering
\begin{tabular}{l|ccccc|ccc|ccc|cccccc}

\toprule

Molecule &
\ce{H2} &
\ce{F2} &
\ce{HCl} &
\ce{LiH} &
\ce{H2O} &
\ce{CH2} &
\ce{O2} &
\ce{BeH2} &
\ce{H2S} &
\ce{NH3} &
\ce{N2} &
\ce{CH4} &
\ce{C2} &
\ce{LiF} &
\ce{PH3} &
\ce{LiCl} &
\ce{Li2O} \\

Valid $\s{k}{}$ &
\num{4} &
\num{100} &
\num{100} &
\num{225} &
\num{441} &
\num{735} &
\num{1.2e3} &
\num{1.2e3} &
\num{3e3} &
\num{3e3} &
\num{14e3} &
\num{16e3} &
\num{44e3} &
\num{44e3} &
\num{48e3} &
\num{1e6} &
\num{41e6} \\

Num. terms ($K$) &
\num{15} &
\num{2951} &
\num{5851} &
\num{631} &
\num{1390} &
\num{2058} &
\num{2879} &
\num{2074} &
\num{9558} &
\num{4929} &
\num{2239} &
\num{8480} &
\num{2239} &
\num{5849} &
\num{24369} &
\num{24255} &
\num{20558} \\

Compute time (\si{\second}) &

\multicolumn{5}{c|}{$\sim 10^{-3}$} &
\multicolumn{3}{c|}{$\sim 10^{-2}$} &
\multicolumn{3}{c|}{$\sim 10^{-1}$} &

\num{1.2} &
\num{1.0} &
\num{2.2} &
\num{7.6} &
\num{192} &
--- \\
	 
\bottomrule

\end{tabular}
\caption{
The number of physically valid determinants, $\s{k}{}$, and Pauli strings in the Jordan-Wigner encoded Hamiltonian, $K$, and the total time to compute all non-zero coupling strengths in the corresponding matrix form of the Hamiltonian, for all molecules considered in this work.  The missing entry for \ce{Li2O} indicates that the required coupling terms were computed on-demand during optimisation.
}
\label{tab:coupling_calcs}
\end{table*}

\subsubsection*{Timing analysis}

\begin{figure}[!h]
\centering
\includegraphics[width=0.9\linewidth]{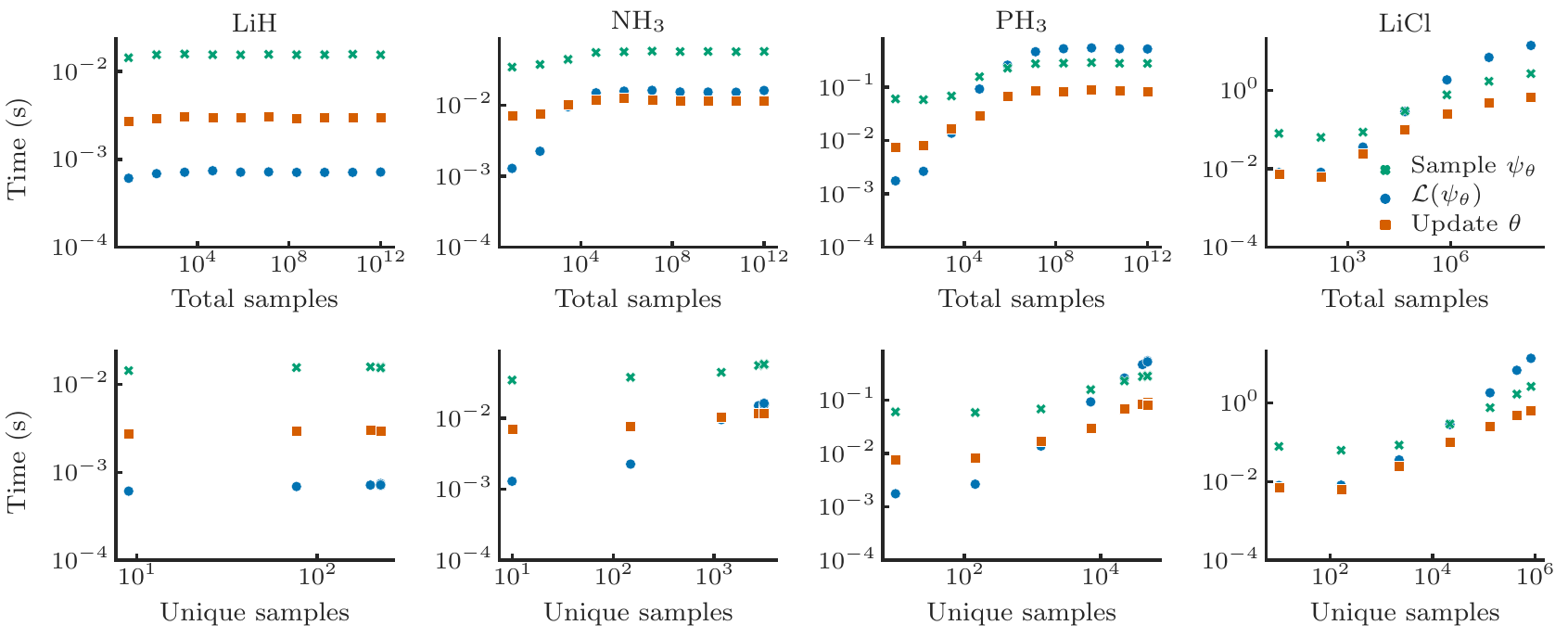}
\caption{
	A breakdown of the time taken for each of the 3 stages of an optimisation step: sampling, calculating the loss and updating the NAQS parameters.  Along the top row, we present these times as a function of the overall batch size (\ie{} number of samples generated), with the bottom row instead using the number of unique samples generated.  Unlike the other molecules plotted, \ce{LiCl} does not go up to $10^{12}$ total samples as the number of unique samples generated becomes too large to fit into the GPU memory.
	As wavefunctions that approach the ground state can have a very narrow distribution, we instead use use randomly initialised NAQS to generate the data to ensure high numbers of unique configurations can be obtained.
}
\label{fig:timings}
\end{figure}

Each optimisation step can be split into three sequential sub-steps -- sampling Slater determinants, evaluating their local energies and updating the wavefunction.  \Cref{fig:timings} summarises the contributions of each of these stages to the overall step time when increasing the number of samples from $10$ to $10^{12}$ on molecules of increasing size.  The top row shows the time taken as a function of the batch size, however, in practice it is the number of unique samples (bottom row) that determines the computational burden.
As we optimise the NAQS, the distribution defined by the wavefunction becomes increasingly focussed on only the relevant Slater determinants, meaning fewer unique samples are generated from a given size of sampled batch.
Therefore, as discussed in the Methods, in practice we tune the batch size -- to an upper limit of $10^{12}$ ($10^{6}$) total (unique) samples -- to keep the total number of unique samples between $10^4$ and $10^5$ for as long as possible.

\vspace{\baselineskip}
\textbf{Sampling.~} The first step is to sample a batch of $B$ Slater determinants and the corresponding complex amplitudes from the wavefunction, $\mathcal{B} = \{ (\s{i}{}, \psi_\theta(\s{i}{}) )_j \sim \abs{\psi_\theta}^{2}) \vert j{=}1,\dots,N \}$.  As discussed in the main text, each sub-network of our NAQS computes the conditional distributions over the possible occupancy states of the next spatial orbital (pair of qubits), $\psi_{i,\theta}(v_k^i \vert v_k^1 \dots v_k^{i-1})$, where $v_k^i \equiv (\se{k}{i\uparrow}, \se{k}{i\downarrow})$ can be one of four possible configurations: $(0,0), (0,1), (1,0), (1,1)$.  This can be done in a single batched forward-pass of the $i$-th sub-network, processing all unique configurations of $v_k^1 \dots v_k^{i-1}$ sampled thus far in parallel.

Sampling from the multinomial distribution output by the sub-network can be performed extremely efficiently, therefore the forward-pass of the network accounts for the bulk of time required to generate a batch of Slater determinants.  For example, consider the case when we have $10^{12}$ samples evenly distributed across $10^5$ unique states (both set to the upper bounds we restrict our sampling to in this work).  Each sub-network would generate up to $10^5$ 4-dimensional discrete distributions for the occupancy of the next spatial orbital.  Sampling the required statistics from these can be performed in parallel, taking only \SI{24.3 \pm 0.8}{\ms} on a standard MacBook (2.3 GHz Quad-Core Intel Core i7 processor).  Concretely, for the smallest sample size on \ce{LiH} and largest sample size on \ce{LiCl} showns in \cref{fig:timings}, the contribution of numerically sampling the networks outputs ranges from \SI{1.5}{\percent} to \SI{2.6}{\percent}, with the rest attributable to network inference.

\vspace{\baselineskip}
\textbf{Evaluating local energy and loss.~} Given a sampled batch of Slater determinants, the next step is to calculating the local energies, $E\sub{loc}{(\s{k}{})}$ (eq.\ (8) of main text), and combine these into an approximation of the wavefunction's energy, \ie{} the `loss' function we want to minimise,
\begin{equation}
	\mathcal{L}(\psi_\theta) = 2\Re \mathbb{E}_{\mathcal{B}} \left[ E\sub{loc}(\psi_\theta)_\bot \ln(\psi_\theta^{\ast}) \right].
\end{equation}
Here, $\bot$ indicates a stop gradient operation and we note that $\nabla_{\theta} \mathcal{L}$ is equivalent to eq.\ (9) of the main text.
Again, the computational cost of this step scales with the number of unique Slater determinants sampled.  Concretely, the computational cost of calculating the local energies of $N\sub{unq}$ unique samples for a system with $\Nq$ single-electron orbitals scales as $\bigO{\Nq^4 N\sub{unq}}$~\cite{mcardle20}.  From \cref{fig:timings}, we see that practically, whilst sampling the NAQS is the dominant computational cost for small molecules, for larger molecules the calculation of the loss given these samples becomes the rate-determining step.  Not included in the presented timings is the cost of calculating the couplings between Slater determinants.  However, for very large systems, such as \ce{Li2O}, pre-calculating all terms of the Hamiltonian is impractical and so they are instead calculated on demand, further increasing the cost of this step.

\vspace{\baselineskip}
\textbf{Update the wavefunction.~} Given the loss, the final step is to calculate and apply an update to the NAQS parameters, $\theta$, using the standard backpropagation algorithm.  Practically, we find that this scales proportionally with the cost of a sampling from the network (\ie{} a network forward-pass plus peripheral processing), but is strictly a cheaper operation.

\vspace{\baselineskip}
\textbf{Overall timings.~} It is not straightforward to consider the breakdown of a single optimisation step presented in \cref{fig:timings} and infer the expected wall-clock time of a full optimisation.  The number of unique samples largely determines the step time, however this is highly non-stationary as the wavefunction distribution $\abs{\psi_\theta}^2$ is updating over the course of training.  For large systems with many unique configurations sampled, the calculation of local energy and the network loss dominate, however smaller systems or highly optimised wavefunctions typically have the network inference limiting the rate of training.
For example, when optimising \ce{Li2O}, the largest molecule we consider, the first step uses $\SI{100}{k}$ samples distributed across $\SI{95}{k}$ unique configurations, and the overall step time (including calculating the coupling matrix elements) is \SI{34}{\second}.  However, by the end of training, $10^{12}$ samples are distributed across $\SI{9}{k}$ unique configurations and the overall step time is down to \SI{2}{\second}.

\begin{figure}[!t]
\centering
\includegraphics[width=0.8\linewidth]{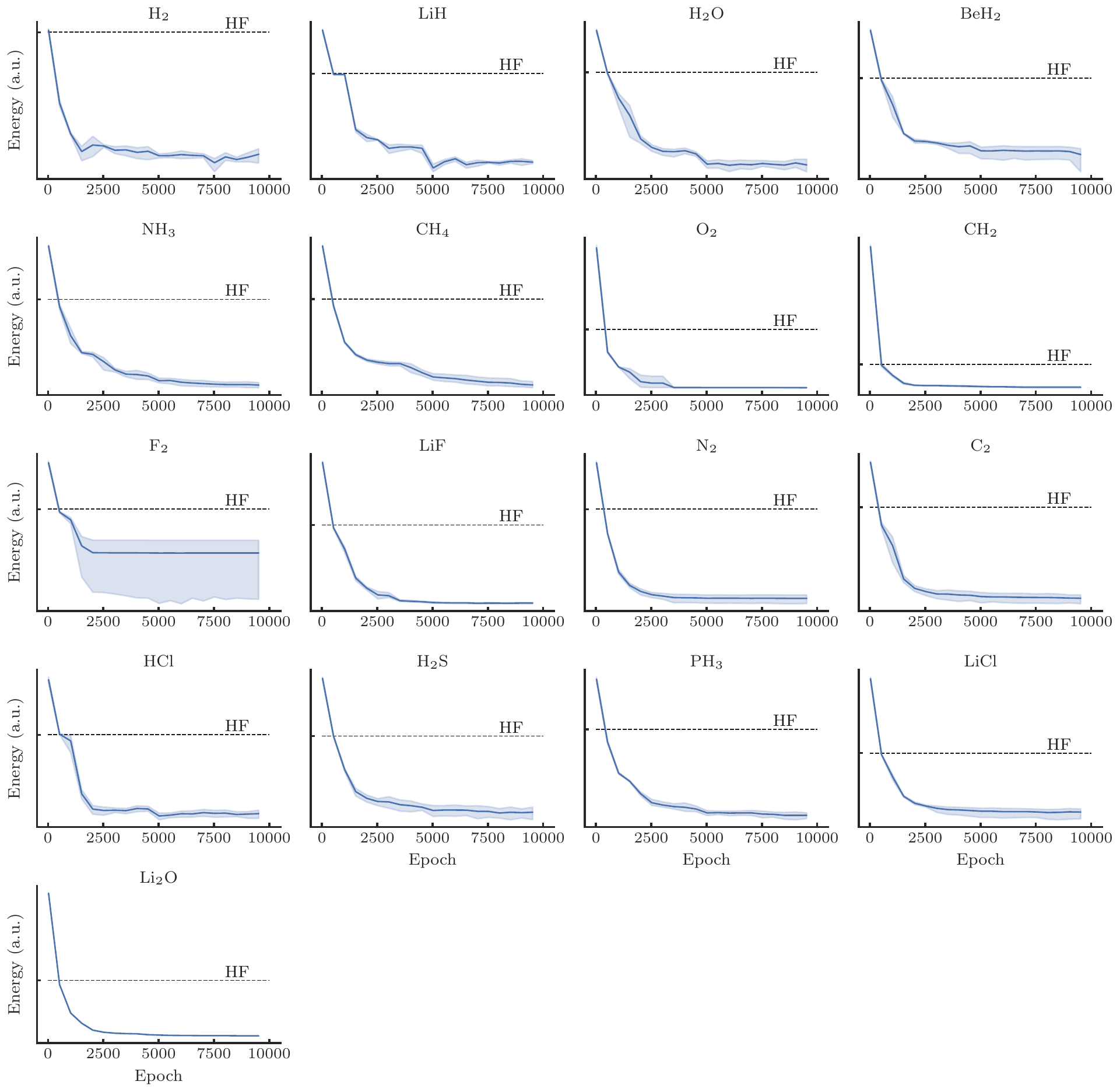}
\caption{The variation energies obtained over the course of optimisation using the standard network architecture presented in the main text.  Where possible, the results are averaged across all five training seeds, with the shaded regions denoting the \SI{95}{\percent} confidence interval.  For \ce{Li2O}, the single optimisation attempt for each network configuration is displayed.  All plots use logarithmic scaling on the energy axis, with a dashed horizontal line denoting the Hartree-Fock (HF) energy.}
\label{fig:allLearningCurves}
\end{figure}

Whilst reasonable effort was made to ensure efficient implementations of the required operations, there remain many paths to further improving speed (both algorithmic, such as avoiding conversions between sparse and dense matrices and technical, where the required combination of neural networks and batched matrix operations would make our NAQS ideal for a compiled auto-grad library such as Jax~\cite{jax18}).  Moreover, the hyperparameters of training are fixed and are were not tuned for fastest performance -- indeed, it is clear from the learning curves in \cref{fig:allLearningCurves} that many molecules are already optimised before the training run ends.  However, to provide context we can still compare the overall wall-time our our NAQS optimisations with classical methods, as presented in \cref{tab:wall_times}.

\begin{table*}[h]
\centering
\begin{tabular}{c cc ccccc}

\toprule

&  \multicolumn{2}{c}{Psi4~\cite{smith20}} & \multicolumn{5}{c}{NAQS} \\

\cmidrule[0.75pt](lr){2-3} \cmidrule[0.75pt](lr){4-8}

Molecule & CCSD(T) & FCI & $0.95E\sub{best}$ & $0.99E\sub{best}$ & $0.995E\sub{best}$ & $0.999E\sub{best}$ & $E\sub{best}$ \\

\cmidrule(lr){1-8}

\ce{LiH} &
\SI{1.5}{\second} & \SI{1.1}{\second} &
\SI{1.68\pm0.14}{\second} & \SI{3.34\pm0.08}{\second} & \SI{6.03\pm0.12}{\second} & \SI{26.5\pm1.7}{\second} & \SI{79.6\pm13.8}{\second} \\

\ce{NH3} &
\SI{2.0}{\second} & \SI{1.1}{\second} &
\SI{5.0\pm1.4}{\second} & \SI{8.3\pm0.7}{\second} & \SI{17.6\pm1.6}{\second} & \SI{36.0\pm3.8}{\second} & \SI{6.1\pm0.7}{\minute} \\

\ce{PH3} &
\SI{3.3}{\second} & \SI{2.4}{\second} &
\SI{31.1\pm1.1}{\second} & \SI{50.9\pm2.8}{\second} & \SI{1.9\pm0.6}{\minute} & \SI{4.2\pm0.3}{\minute} & \SI{17.5\pm2.5}{\minute} \\

\ce{LiCl} &
\SI{5.0}{\second}$^{\ast}$ & \SI{35.8}{\second} &
\SI{1.3\pm0.1}{\minute} & \SI{2.3\pm0.1}{\minute} & \SI{3.1\pm0.2}{\minute} & \SI{2.8\pm0.1}{\hour} & \SI{1.5\pm0.2}{\hour} \\

\ce{Li2O} &
\SI{8.0}{\second}$^{\ast}$ & \SI{58.8}{\minute} &
\SI{13.1}{\minute} & \SI{21.5}{\minute} & \SI{30.5}{\minute} & \SI{4.1}{\hour} & \SI{1.9}{\day} \\

\bottomrule

\end{tabular}
\caption{
Wall-clock times for solving a selection of molecules using classical methods and NAQS.  As the NAQS optimisation does not have a termination criterion (other than the \SI{10}{k} steps we set), we present the time taken to reach various fractions of the best energy found by the NAQS, $E\sub{best}$.  The $^{\ast}$ denotes that CCSD(T) returns non-physical energies (that are lower than the ground-truth FCI energy) on the largest molecules (see Table 1 of the main text for numerical results).
}
\label{tab:wall_times}
\end{table*}

\end{document}